\documentclass[traditabstract]{aa}  

\usepackage{natbib}
\usepackage[utf8]{inputenc}
\usepackage{graphicx}
\usepackage[varg]{txfonts}
\usepackage{amsmath}
\usepackage{mathtools}
\usepackage[usenames,dvipsnames]{xcolor}
\usepackage{subcaption}
\newcommand{\pb}[1]{{#1}}

\newcommand{\isrev}[1]{{#1}}

\begin{document}

   \title{Supernova remnants in super bubbles as cosmic ray accelerators}

   \author{Iurii Sushch
          \inst{1,2,3,4,5}\thanks{\email{iurii.sushch@ciemat.es}}
          \and
          Pasquale Blasi\inst{2,3}
          \and
          Robert Brose\inst{6,7}
          }

   \institute{ Centro de Investigaciones Energ\'eticas, Medioambientales y Tecnol\'ogicas (CIEMAT), E-28040 Madrid, Spain 
        \and
        Gran Sasso Science Institute, Via F. Crispi 7, 67100       L’Aquila, Italy
        \and
        INFN-Laboratori Nazionali del Gran Sasso, Via G. Acitelli 22, Assergi (AQ), Italy
        \and
        Centre for Space Research, North-West University, 2520 Potchefstroom, South Africa
        \and
        Astronomical Observatory of Ivan Franko National University of Lviv, Kyryla i Methodia 8, 79005 Lviv, Ukraine
        \and
        Institute of Physics and Astronomy, University of Potsdam, 14476 Potsdam-Golm, Germany
        \and
        School of Physical Sciences and Centre for Astrophysics \& Relativity, Dublin City University, Glasnevin, D09 W6Y4, Ireland.
}

   \date{Received ; accepted }

 
  \abstract
  {Supernova remnants (SNRs) are often considered as the main sites of acceleration of cosmic rays in our Galaxy, possibly up to the knee. However, their ability to accelerate particles to reach PeV energies is questionable and lacks observational evidence. Theoretical predictions suggest that only a small subclass of very young SNRs evolving in dense environments could potentially satisfy the necessary conditions to accelerate particles to PeV energies. Most such theoretical investigations are carried out either in the standard interstellar medium or in the wind of the progenitor. Since most core collapse supernovae occur in star clusters, it is important to extend such investigation to SNRs taking place in a star cluster.
  In this work we focus on a SNR shock propagating in the collective wind of a compact star cluster, and we study the acceleration process as a function time, with special emphasis on the maximum energy of accelerated particles. Using both analytic and numerical approaches we investigate the spectrum of accelerated particles and maximum achievable energy in the case of pre-existing turbulence in the collective wind and self-generated magnetic perturbations. We find that similar to isolated SNRs, acceleration to PeV energies is plausible only for extreme conditions achievable only in a small subset of SNRs.}


   \maketitle
%

\section{Introduction}

For a long time supernova remnants (SNRs) have been considered as the main candidates to being sources of the bulk of Galactic cosmic rays (CRs), mainly on the basis of energetic arguments. Gamma-ray observations provide an indirect proof that SNRs can accelerate particles to energies as high as $\sim 10-100$ TeV \citep{2013Sci...339..807A,2011MmSAI..82..747G,2016ApJ...816..100J,2019A&A...623A..86A, 2020MNRAS.497.3581D} implying that SNRs certainly are important contributors to the CR sea at low energies. However, the shape of the CR spectrum, in particular the distinct 'knee' spectral feature at $\sim3$~PeV, strongly suggests that protons should be accelerated at least to PeV energies within our Galaxy. Moreover, if the transition from Galactic to extragalactic CRs occurs in fact at $\sim 10^{17}$ eV, at the less pronounced 'second knee' in the CR spectrum, then Galactic sources should be able to accelerate protons beyond 10 PeV \citep[see][and references therein]{2023MNRAS.519..136V}. Besides the lack of observational evidence for SNRs as PeVatrons, there is also a growing theoretical consensus that only exceptionally energetic SNRs \isrev{or very young SNRs (tens of years after explosion) in specific environments} might accelerate protons to the knee region \citep{2018MNRAS.479.4470M, 2020APh...12302492C, 2020MNRAS.494.2760C, 2021ApJ...922....7I, 2022MNRAS.516..492B, 2023ApJ...958....3D, 2025arXiv250420601B}.

A number of other source types have been proposed to take over at PeV energies, among which the most promising in the light of recent observations seem to be young compact star clusters \citep{2019NatAs...3..561A, 2021NatAs...5..465A,2022A&A...666A.124A, 2024SciBu..69..449L} and microquasars \citep{2024Sci...383..402H, nature.hawc.V4641Sgr, 2024arXiv241008988L}. The former one is interesting in the context of the SNR paradigm, because one of the scenarios considered for particle acceleration in these objects are SNRs evolving in the bubble generated by the collective wind from the massive stars in the cluster. \citet{2022MNRAS.515.2256V} proposed that the SNR shock expanding into the collective wind with pre-existing magnetic turbulence advected from the cluster could accelerate particles to $\sim10$\,PeV without the need for additional self-generated magnetic field amplification. This scenario is attractive also because it seems to be able to accommodate the composition of CRs which is different from the solar one. In particular, the ratio of $^{22}$Ne to $^{20}$Ne, which is $\sim5$ times higher than the solar value \citep[see e.g.][]{2008NewAR..52..427B} can be explained by the enrichment of the super bubble environment with Wolf-Rayet winds \citep{2020MNRAS.493.3159G}. 
\isrev{The majority of massive stars reside in clusters strongly suggesting that most of core-collapse supernovae occur in stellar clusters \citep{2004A&A...424..747P, 2010ARA&A..48..431P}. Unambiguous observational evidence is lacking mainly due to the fact that supernovae are not particularly frequent events and the capabilities of instruments to detect star cluster counterparts are limited to nearby galaxies \citep{2010ARA&A..48..431P}. One of the most prominent associations reported is the type IIp supernova SN2004dj which is believed to occur in the star cluster Sandage-96 in the spiral galaxy NGC 2403 \citep{2005ApJ...626L..89W}. Nevertheless, if this is the case, it is expected that a large fraction of SNRs are evolving in super bubbles with an environment potentially favourable for effective particle acceleration. However, no detailed simulations have been available so far that would quantitatively confirm acceleration of particles to PeV energies in such an environment.}



There are several potential sites of particle acceleration \isrev{and subsequent non-thermal emission in super bubbles \citep[see e.g.][]{2004A&A...424..747P} and existing observations do not offer an unambiguous preference to anyone of them. Particles can be accelerated inside cores of stellar clusters through shock re-acceleration, shock collisions, and stochastic re-acceleration, but recent simulations show that these processes are ineffective in boosting the maximum energy of accelerated particles \citep{2024MNRAS.530.4747V}. There is also no observational evidence of high-energy emission from the cores. Some clusters like Westerlund 1 show a shell-like morphology of the associated gamma-ray emission \citep{2022A&A...666A.124A} and for those that exhibit centrally filled morphology like Cygnus OB2 the angular resolution of instruments does not allow for firm identification of the emission with the core \citep{2019NatAs...3..561A}.}

\isrev{Another potential site of the of effective particle acceleration which also has some observational support is the termination shock of the collective wind of the stellar cluster. For instance, H.E.S.S. observations of Westerlund 1 show the shell-like morphology of the gamma-ray emission with the radial distance to the peak of the gamma-ray emission of $\sim34$~pc that agrees well with the estimate of the radius of the termination shock \citep{2022A&A...666A.124A}.} \pb{In this framework, \cite{Westerlund1-2023} advocated for a leptonic scenario where most gamma ray emission is due to ICS of high energy electrons accelerated at the termination shock, while the contribution of hadronic interactions was estimated to be negligible under standard assumptions. It remains to be seen if this scenario is compatible with the absence of radio emission and with the morphology of the gamma ray emission.}
\pb{From the theoretical point of view, \citet{2021MNRAS.504.6096M} showed that acceleration at the termination shock}
can be efficient, but PeV energies can be reached only for rather bright star clusters with very fast outflows and only for Bohm diffusion, while for Kolmogorov and Kraichnan diffusion the effective maximum energy falls short of the 1 PeV mark. 

\isrev{It is also possible that particles could be accelerated in the super bubble downstream of the termination shock through the second order Fermi acceleration. It is argued that a high level of turbulence and high Alfv\'{e}n velocity could significantly speed up this acceleration mechanism, but, similar to the acceleration at the termination shock even for optimistic considerations PeV energies are achievable only for Bohm diffusion \citep{2024arXiv240603555V}, which is a rather unrealistic description of the turbulence in the super bubble.}

In the following we \isrev{analyse in depth} the remaining scenario that is linked to the evolution of the putative SNR exploded at the edge of the star cluster core and expanding into the turbulent magnetized environment of the star cluster super bubble. \isrev{The main objective of the study is to critically asses the ability of an SNR evolving in a super bubble to accelerate particles to energies beyond the knee. We consider the most favourable scenario whereas the putative SNR is the remnant of the first supernova explosion in the star cluster and the star cluster is compact enough to produce a collective wind. It is also expected that the most effective acceleration would happen at a relatively young age of an SNR before the forward shock reaches the termination shock and slows down considerably. It should be noted that although a supernova explosion in a young massive stellar cluster is unavoidable and its occurrence at the edge of a cluster is quite probable, there is very limited observational evidence of SNRs identified within super bubbles, especially for the cases where efficient particle acceleration is confirmed through X-ray and/or gamma-ray observations. This is partly due to the difficulties of simultaneous detection of both objects and reliable association with each other, and partly due to the selection bias linked to the fact that the whole lifetime of an SNR is only a small fraction of the lifespan of the stellar cluster. Nevertheless, in some cases there are convincing arguments that indeed we witness the evolution of an SNR inside the super bubble generated by the stellar cluster. For example, X-ray emission associated with the super bubble 30 Dor C in the Large Magellanic Cloud requires high shock velocities, which strongly suggests that this emission is generated by an SNR within the super bubble \citep{2004ApJ...602..257B, 2009PASJ...61S.175Y, 2015Sci...347..406H}. The picture depicts, however, late stages of the evolution of the SNR when it starts to interact with the outer shell of the super bubble.}

The paper is structured as follows: in Section \ref{sec:coll_wind}, we describe the environment in which the remnant evolves, in Section \ref{sec:Emax} we provide analytic calculations of the maximum energy based on different assumptions for the magnetic turbulence responsible for the confinement of particles at the shock, in Section \ref{sec:numsim} we describe the numerical setup and present the results on numeric simulations using the RATPaC software, and in Section \ref{sec:summary} we summarize and discuss our findings. 




\section{The ambient medium}
\subsection{Collective wind}
\label{sec:coll_wind}
\begin{figure}
    \centering
    \includegraphics[width=\linewidth]{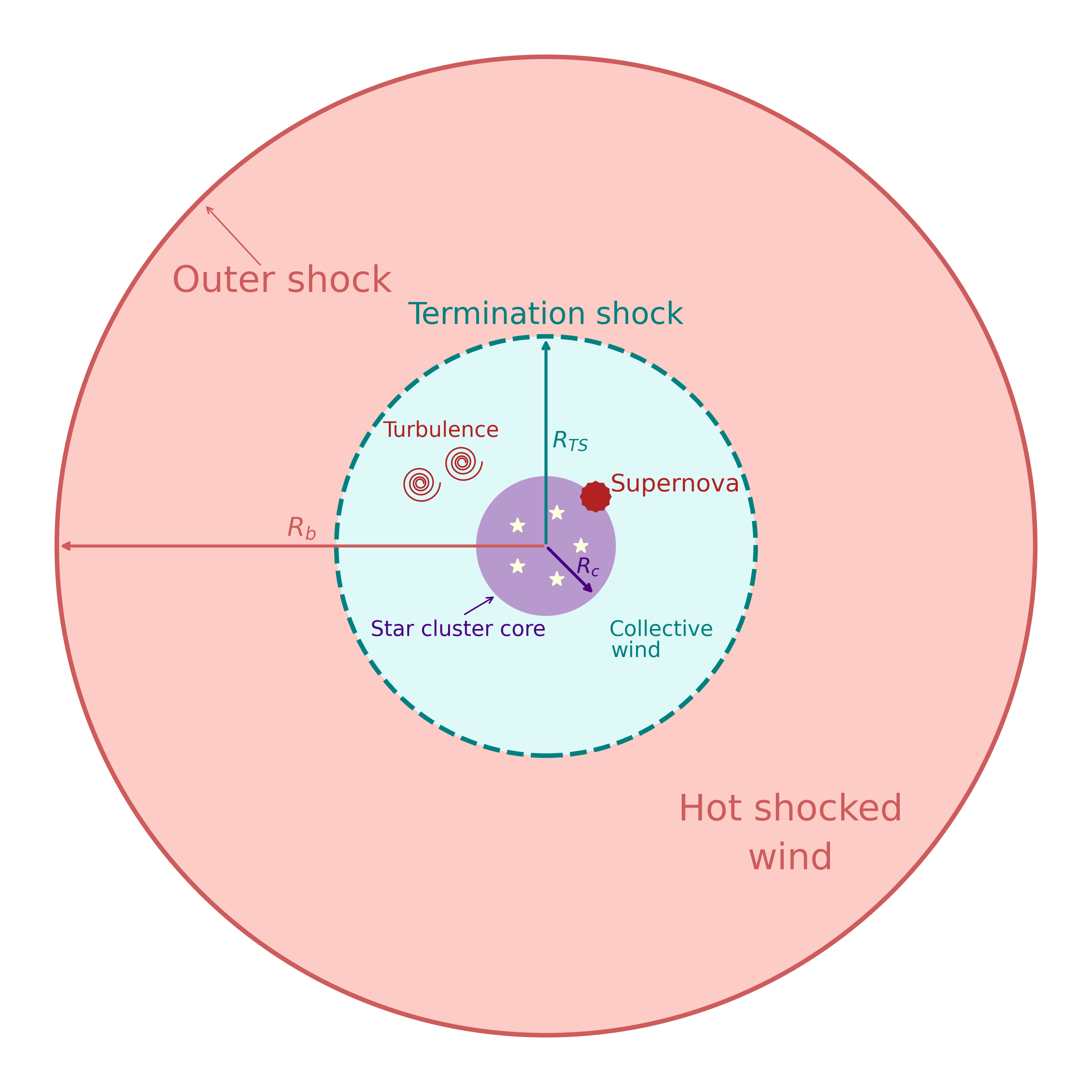}
    \caption{\isrev{A schematic cartoon illustrating the considered scenario. Note, ratios of the relevant length scales, i.e. the radius of the core, $R_\mathrm{c}$, the radius of the termination shock, $R_\mathrm{TS}$, and the radius of the outer shock, $R_\mathrm{b}$, do not reflect the real proportions.}}
    \label{fig:cartoon}
\end{figure}

In the following we consider a supernova exploding at the edge of the core of the young compact stellar cluster in which we assume that no previous explosions occurred \isrev{(see Fig. \ref{fig:cartoon} for a schematic cartoon depicting the scenario considered)}. The SNR shock expands into the medium shaped by the collective wind of the stars in the stelalr cluster. Typically, stellar clusters contain under 100 supernova progenitors \citep{2018AdSpR..62.2750L}, while the most massive stellar clusters may contain up to several thousand massive stars \citep[see e.g.][]{2019ARA&A..57..227K}. \isrev{The most massive stellar cluster in Milky Way, Westerlund 1, hosts at least 166 stars with initial masses between $\sim25 M_\odot$ $\sim50 M_\odot$ out of which 24 are classified as Wolf-Rayet stars \citep{2020A&A...635A.187C}.}
The first supernova explosion takes place at the age of around $3$ million years. The average mechanical power of the collective wind of the cluster consisting of 100 stars throughout its evolution is of the order of $\sim10^{38}$~erg/s \citep{2022MNRAS.512.1275V}. In the following we consider the total mass-loss rate of $\dot{M}=2\times10^{-4}\,M_\odot/\mathrm{yr}$ and wind velocity of $v_\mathrm{w}=3\times10^8$~cm/s, that is equivalent to 100 typical Wolf-Rayet or very luminous main sequence stars and corresponds to the total wind power of $L_\mathrm{w}=\frac{1}{2}\dot{M}v_\mathrm{w}^2=6\times10^{38}$~erg/s. The choice of parameters is identical to the Model 4 considered for Cygnus OB2 in \citet{2023MNRAS.523.4015B}. The location of the termination shock can be approximated by \citep{2021MNRAS.504.6096M, 2023MNRAS.523.4015B}
\begin{multline}
    R_\mathrm{TS} \simeq 21 \left[\frac{\dot{M}}{2\times10^{-4}\,M_\odot/\mathrm{yr}} \right]^{3/10}\left[\frac{n_\mathrm{ISM}}{10\,\mathrm{cm}^{-3}}\right]^{-3/10} \times \\
    \left[\frac{v_\mathrm{w}}{3\times10^8\,\mathrm{cm/s}}\right]^{1/10} \left[\frac{t}{3\times10^6\,yr}\right]^{2/5}\,\mathrm{pc}    
\end{multline}
which agrees within $\lesssim 10\%$ with a more accurate \citet{1977ApJ...218..377W} calculation. The position of the outer shock of the bubble can be approximated by
\begin{multline}
    R_\mathrm{b} \simeq 120 \left[\frac{\dot{M}}{2\times10^{-4}\,M_\odot/\mathrm{yr}} \right]^{1/5}\left[\frac{n_\mathrm{ISM}}{10\,\mathrm{cm}^{-3}}\right]^{-1/5} \times \\
    \left[\frac{v_\mathrm{w}}{3\times10^8\,\mathrm{cm/s}}\right]^{2/5} \left[\frac{t}{3\times10^6\,yr}\right]^{3/5}\,\mathrm{pc}.    
\end{multline}

The gas density in the wind region can be written as 
\begin{equation}
\label{eq:rho}
    \rho = \frac{\dot{M}}{4\pi r^2 v_\mathrm{w}},
\end{equation}
up to the termination shock. Downstream of the shock the density is assumed to be constant while the velocity of the shocked wind drops as $\propto r^{-2}$. These trends are to be considered as valid descriptions of the wind for compact stellar cluster, while non compact clusters might have a complex outflow that however cannot be described in terms of a collective wind. In fact, this might be the case of Cygnus OB2, where 3D hydrodynamic simulations based on the observed stellar population suggest that no collective wind is formed \citep{2024MNRAS.532.2174V}.


 \subsection{Magnetic field and pre-existing turbulence}
A fraction of the kinetic energy of the winds of individual stars is expected to be converted into magnetic turbulence through wind-wind collisions. This turbulence is advected with the collective wind and compressed at the termination shock. It is however unclear how much of the total kinetic energy of winds is dissipated \isrev{and what is the level of magnetic turbulence}. \isrev{Numerical 3D MHD simulations show that the magnetic field in the stellar cluster core is non-uniform reaching values of 1~$\mu$G with the mean value of the magnetic field reaching $200\,\mu$G and median value reaching $35\mu$G \citep{2025arXiv250319745H}. }

It should be understood that the collective wind parameters set above already account for part of the total kinetic energy being converted into turbulence. Assuming passive transport of the turbulent magnetic field with the flow and ignoring any potential damping, the turbulent field can be written as
\begin{equation}
\label{bturb}
    B(r) = 
    \begin{dcases}
        B_\mathrm{c} \frac{R_\mathrm{c}}{r}, & r\leq R_{\mathrm{TS}},\\
        \sqrt{11}  B_\mathrm{c}\frac{R_\mathrm{c}}{R_{\mathrm{TS}}}, & R_{\mathrm{TS}} < r < R_{\mathrm{b}}
    \end{dcases}
\end{equation}
where $B_\mathrm{c}$ is the generated turbulent field at the edge of the cluster core at the radius of $R_\mathrm{c}$ and $\sqrt{11}$ is the compression factor at the termination shock. \isrev{The 3D MHD simulations presented in \citep{2025arXiv250319745H} confirm tha the magnetic field is indeed scaled as approximately $1/r$ in the collective wind.} The magnetic field is assumed to stay constant in the bubble downstream of the termination shock, which should be considered as an optimistic scenario, because the turbulent field could be subjected to significant damping.
The total energy in the turbulent magnetic field can be estimated as 
\begin{equation}
    E_\mathrm{turb} = \int_{R_\mathrm{c}}^{R_\mathrm{b}} \frac{B(r)^2}{8\pi} 4\pi r^2dr
\end{equation}
This can be compared to the total energy in the wind calculated as 
\begin{equation}
    E_\mathrm{wind} = 5.7\times10^{52} \left[\frac{L_\mathrm{w}}{6\times10^{38}\,\mathrm{erg/s}}\right]\left[\frac{t}{3\times10^6\,yr}\right]\,\mathrm{erg}
\end{equation}
Setting $E_\mathrm{turb} = \eta_\mathrm{E} (E_\mathrm{wind}+E_\mathrm{turb})$ and assuming $R_\mathrm{c} = 1$~pc one can estimate the the strength of the turbulent magnetic field at the edge of the core for adopted parameters of the wind as 
\begin{equation}
    B_\mathrm{c}= 178  \,\mu\mathrm{G}
\end{equation}
for $\eta_\mathrm{E}=0.1$

Alternatively, one could constrain $B_\mathrm{c}$ by assuming that the magnetic energy density at the edge of the core constitutes a fraction of the wind energy
\begin{equation}
    \frac{B_\mathrm{c}^2}{4\pi} = \frac{\eta_\mathrm{w}}{1-\eta_\mathrm{w}} \rho_\mathrm{w} v_\mathrm{w}^2 = \frac{\eta_\mathrm{w}}{1-\eta_\mathrm{w}}\frac{\dot{M}v_\mathrm{w}}{4\pi}
\end{equation}
For assumed parameters of the collective wind the turbulent magnetic field at the edge of the core for $\eta_\mathrm{w}=0.1$ is 
\begin{equation}
    B_\mathrm{c}= 222 \,\mu\mathrm{G}.
\end{equation}
Since this magnetic field is assumed to be created through the dissipation of the winds of the many stars in the core, it is expected to be turbulent, with some coherence scale and that it eventually cascades to smaller scales. \isrev{These estimates are in agreement with the mean values obtained in 3D MHD simulations but they are higher than the bulk magnetic field strength represented by the median values \citep{2025arXiv250319745H}. For the radius of the termination shock of $\sim20$~pc (see Section~\ref{sec:coll_wind}) the estimated value of $B_\mathrm{c}\sim200\,\mu$G corresponds to $\sim10\,\mu$G at the termination shock and to $\sim30\,\mu$G in the super bubble downstream of the termination shock. Available measurements suggest lower values of the magnetic field: $B\approx 3-10\,\mu$G in Orion-Eridanus \citep{2019A&A...631A..52J} and $B\approx 10-20\,\mu$G in 30 Dor C \citep{2019A&A...621A.138K}. In this context adopting in the following $B=10\,\mu$G in the immediate upstream of the termination shock we explore the high end of the parameter space allowing for as high maximum energies as possible.}

\section{Estimates of the maximum energy}
\label{sec:Emax}
\subsection{Pre-existing turbulence}
We assume that the turbulence is generated at the coherence scale of $L=1$~pc that corresponds to the assumed radius of the core. Further, we assume that the turbulence evolves following either the Kolmogorov or Kraichnan cascade, hereafter referred as KOLM and KRAI. The diffusion coefficient for these two cases can be expressed as 
\begin{equation}
\label{eq:Dkolm}
D_\mathrm{KOLM} = \frac{1}{3}r_\mathrm{L}v\left(\frac{r_\mathrm{L}}{L}\right)^{-2/3}
\end{equation}
\begin{equation}
\label{eq:Dkrai}
D_\mathrm{KRAI} = \frac{1}{3}r_\mathrm{L}v\left(\frac{r_\mathrm{L}}{L}\right)^{-1/2}
\end{equation}
where $r_\mathrm{L} = pc/(eB)$ is the Larmor radius of the particle with momentum $p$ and $v$ is the particle velocity.

The maximum energy that can be achieved in the acceleration process can be estimated limiting the acceleration time with the age of the remnant. The energy gain rate of the particle can be written as 
\begin{equation}
\frac{dE}{dt} \simeq \frac{\Delta E}{\tau_\mathrm{cycle}} = \frac{u_1 - u_2}{3}\left(\frac{D_1}{u_1} + \frac{D_2}{u_2}\right)^{-1} E,   
\end{equation}
where $\Delta E = \frac{4}{3}\frac{u_1-u_2}{c} E$ is the energy gain per one acceleration cycle and $\tau_\mathrm{cycle} = \frac{4}{c}\left(\frac{D_1}{u_1} + \frac{D_2}{u_2}\right)$ is the cycle time interval. Upstream and downstream regions are denoted with subscripts $1$ and $2$ respectively with $u$ being the plasma velocity and $D$ the diffusion coefficient. For $u_1 = v_\mathrm{s}$ and assuming a strong shock with the compression ratio of $u_1/u_2 = 4$ this can be further rewritten as
\begin{equation}
    \label{eq:dEdt}
    \frac{dE}{dt} = \frac{v_\mathrm{s}^2}{4 (D_1+4D_2)}E
\end{equation}

For fully isotropic turbulence the magnetic field downstream is compressed by a factor of $B_2 = \sqrt{11}B_1$, because the radial component of the field is not compressed. Taking into account the dependence of the diffusion coefficient on the magnetic field (see Eqs.~\ref{eq:Dkolm} and \ref{eq:Dkrai}) the relation of the downstream to upstream diffusion coefficient for the two cases of the turbulence spectrum can be estimated as
\begin{equation}
    D_\mathrm{KOLM,2}=0.67D_\mathrm{KOLM,1}\equiv0.67D_\mathrm{KOLM},
\end{equation}
\begin{equation}
    D_\mathrm{KRAI,2}=0.55D_\mathrm{KRAI,1}\equiv0.55D_\mathrm{KRAI}.
\end{equation}
Then assuming that the evolution of the SNR shock is given by 
\begin{equation}
    r_\mathrm{s}(t)\propto t^{j},\,v_\mathrm{s}(t)\propto t^{j-1}
\end{equation}
and 
\begin{equation}
    \label{eq:bfieldt}
    B \propto 1/r
\end{equation}
the maximum energy for the two cases can be estimated by integration of Eq. \ref{eq:dEdt} as
\begin{multline}
    \label{eq:EmaxKOLM}
    E_\mathrm{max}^\mathrm{KOLM} = 23 \left(\frac{j}{5j-3}\right)^3 \left[\frac{t}{t_\mathrm{TS}}\right]^{5j-3} \left[\frac{L}{1\,\mathrm{pc}}\right]^{-2} \left[\frac{R_\mathrm{TS}}{20\,\mathrm{pc}}\right]^3 \\
    \left[\frac{v_\mathrm{TS}}{10^9\,\mathrm{cm/s}}\right]^{3} \left[\frac{B_\mathrm{TS}}{10\,\mu\mathrm{G}}\right]\,\mathrm{TeV}
\end{multline}
\begin{multline}
    \label{eq:EmaxKRAI}
    E_\mathrm{max}^\mathrm{KRAI} = 227 \left(\frac{j}{3j-2}\right)^2 \left[\frac{t}{t_\mathrm{TS}}\right]^{3j-2} \left[\frac{L}{1\,\mathrm{pc}}\right]^{-1} \left[\frac{R_\mathrm{TS}}{20\,\mathrm{pc}}\right]^2 \\
    \left[\frac{v_\mathrm{TS}}{10^9\,\mathrm{cm/s}}\right]^{2} \left[\frac{B_\mathrm{TS}}{10\,\mu\mathrm{G}}\right]\,\mathrm{TeV}
\end{multline}
normalizing the parameters to the values at the time $t_\mathrm{TS}$ when the shock reaches the termination shock with $R_\mathrm{TS}$ and $v_\mathrm{TS}$ being the shock radius and velocity at that time and $B_\mathrm{TS}$ the magnetic field upstream the termination shock. For a core collapse SNR evolving in the free wind at the the free-expansion stage 
$j=6/7$ is typically assumed\footnote{$j=(n-3)/(n-s)$, where $n$ is the power-law index of the ejecta profile, which is commonly assumed to be 9 for core-collapse SNRs and $s$ is power-law index of the density profile, $s=2$ for the wind zone.}. Adopting this value Eqs. \ref{eq:EmaxKOLM} and \ref{eq:EmaxKRAI} can be rewritten as

\begin{multline}
    \label{eq:EmaxKOLM}
    E_\mathrm{max}^\mathrm{KOLM} = 7 \,\,\mathrm{TeV} \\
    \left[\frac{t}{t_\mathrm{TS}}\right]^{9/7} \left[\frac{L}{1\,\mathrm{pc}}\right]^{-2} \left[\frac{R_\mathrm{TS}}{20\,\mathrm{pc}}\right]^3 
    \left[\frac{v_\mathrm{TS}}{10^9\,\mathrm{cm/s}}\right]^{3} \left[\frac{B_\mathrm{TS}}{10\,\mu\mathrm{G}}\right]
\end{multline}
\begin{multline}
    \label{eq:EmaxKRAI}
    E_\mathrm{max}^\mathrm{KRAI} = 510 \,\,\mathrm{TeV} \\
    \left[\frac{t}{t_\mathrm{TS}}\right]^{4/7} \left[\frac{L}{1\,\mathrm{pc}}\right]^{-1} \left[\frac{R_\mathrm{TS}}{20\,\mathrm{pc}}\right]^2 
    \left[\frac{v_\mathrm{TS}}{10^9\,\mathrm{cm/s}}\right]^{2} \left[\frac{B_\mathrm{TS}}{10\,\mu\mathrm{G}}\right]
\end{multline}

In both cases the maximum energy increases with time until the SN shock reaches the termination shock. reaching its peak at the termination shock. At this point, although it may happen that the SN shock is not yet in Sedov phase at this time, the shock velocity drops considerably due to the higher density downstream of the termination shock, so that eventually the Sedov phase is entered shortly after. Furthermore, the high pressure and temperature of the shocked wind would results in a considerable reduction of the shock Mach number and a correspondingly smaller compression ratio at the SN shock. This combination of events causes a softer spectrum of freshly accelerated particles, so that even if the maximum energy may slightly increase, it does not show in the overall spectrum of accelerated particles.

One last point to keep in mind is that in the pre-existing turbulence, the maximum energy is also limited by the largest scale of the turbulence, i.e. coherence scale $L$. Particles with gyroradius larger than $L$ perform small deflection transport so that their maximum energy cannot increase further. This additional criterion leads to the following constraint on the maximum energy \cite[]{2023MNRAS.523.4015B}: 
\begin{equation}
    E_\mathrm{max} = eBL = 9.2 \left[\frac{B_\mathrm{TS}}{10\,\mu\mathrm{G}}\right]  \left[\frac{L}{1\,\mathrm{pc}}\right]\,\mathrm{PeV}.
\end{equation}
This constraint implies that even decreasing the coherence scale in Eqs. \ref{eq:EmaxKOLM} and \ref{eq:EmaxKRAI}, this does not necessarily lead to a higher maximum energy. 


\subsection{Self-generated turbulence}
\label{sec:selfturb}
It is suggestive from the previous section that pre-existing turbulence alone is not sufficient to accelerate particles to PeV energies at the SNR shock. Hence here we consider the possibility that the accelerated particles may drive additional self-generated perturbations through excitation of resonant and non-resonant streaming instabilities, which might help reaching higher maximum energies.

In the bubble excavated by the wind it is hard to imagine the existence of any regular magnetic fields, although one could consider a local direction in which the field is roughly well behaved on a scale of order $\sim L$, the coherence scale of the turbulent field introduced above. On such scales the local CR gradients can excite streaming instability. Given the many assumptions adopted here, it is fair to take the results below as some sort of estimates of the effects expected due to the excitation of instabilities. 

The diffusion coefficient written in quasi-linear theory reads \citep{1978MNRAS.182..147B}
\begin{equation}
\label{eq:Dbohm}
    D = \frac{4}{3\pi} r_\mathrm{L} v \frac{1}{\mathcal{F}}=\eta_\mathrm{B} D_\mathrm{B}
\end{equation}
where $\mathcal{F}$ is the energy density of Alfven waves per unit logarithmic bandwidth scaled to the energy density of the background magnetic field and we introduced a parameter $\eta_\mathrm{B}\equiv 1/\mathcal{F} \geq 1$ characterizing deviations from pure Bohm diffusion. The resonant streaming instability operates on a time scale that is the crossing time of the precursor of particles with momentum $p$ at the shock speed, which results in the following estimate for the $\mathcal{F}$, for the canonical $p^{-4}$ spectrum of accelerated particles \citep{2006MNRAS.371.1251A}:
\begin{equation}
    \mathcal{F} = \frac{1}{\Lambda} \frac{\delta B^2}{B^2}, 
\end{equation}
where $\Lambda = \ln(p_\mathrm{max}/m_\mathrm{p}c) \sim 10$. The corresponding magnetic field strength induced by the instability is \citep{2006MNRAS.371.1251A}
\begin{equation}
    \frac{\delta B^2}{B^2} = 2\frac{v_\mathrm{s}}{v_\mathrm{A}}\eta,
\end{equation}
where $v_\mathrm{A} = B/\sqrt{4\pi\rho}$ is the local Alfv\'en velocity and $\eta$ is the particle injection efficiency defined as the fraction of the ram pressure $\rho v_\mathrm{s}^2$ converted to CRs: $\eta = P_\mathrm{CR}/(\rho v_\mathrm{s}^2)$. 

Combining the equations above and using the expression (\ref{eq:rho}) for the density in the collective wind one can write $\eta_\mathrm{B}$ as
\begin{equation}
    \eta_\mathrm{B} = \frac{1}{2} \frac{Br}{v_\mathrm{s}}\left(\frac{V_\mathrm{w}}{\dot{M}}\right)^{1/2}\frac{\Lambda}{\eta}
\end{equation}
and taking into account evolution of the shock and spatial dependence of the magnetic field in the collective wind
\begin{equation}
    \eta_\mathrm{B} = \frac{1}{2} \frac{B_\mathrm{TS}r_\mathrm{TS}}{v_\mathrm{TS}} \left(\frac{t}{t_\mathrm{TS}}\right)^{1-j}\left(\frac{V_\mathrm{w}}{\dot{M}}\right)^{1/2}\frac{\Lambda}{\eta}.
\end{equation}
For the parameters considered here $\eta_\mathrm{B}$ can be estimated as 
\begin{multline}
     \eta_\mathrm{B} = 4.8 \left[\frac{\eta}{0.1}\right]^{-1}\left[\frac{\Lambda}{10}\right] \left[\frac{\dot{M}}{2\times10^{-4}\,M_\odot/\mathrm{yr}}\right]^{-1/2}\\
     \left[\frac{v_\mathrm{w}}{3\times10^8\,\mathrm{cm/s}}\right]^{1/2} \left[\frac{v_\mathrm{TS}}{10^9\,\mathrm{cm/s}}\right]^{-1}\left[\frac{R_\mathrm{TS}}{20\,\mathrm{pc}}\right] \left[\frac{B_\mathrm{TS}}{10 \mu\mathrm{G}}\right]\left[\frac{t}{t_\mathrm{TS}}\right]^{1-j} 
\end{multline}

A note of caution is in order: although typically $\eta_B\gtrsim 1$, one can see that $\delta B \sim B$, which stretches the validity of this treatment to its limits. The resonant streaming instability is expected to be saturated when $\delta B/B\sim 1$, as a consequence of the requirement that a resonance between waves and particle gyration is fulfilled. When this happens, the most reasonable option is to assume that the diffusion coefficient becomes Bohm-like with $\delta B = B$. With these limitations, one can follow the same line of though outlined above and write
\begin{equation}
    \label{eq:dEdtBohm}
    \frac{dE}{dt} = \frac{v_\mathrm{s}^2}{20\eta_\mathrm{B}D_\mathrm{B}}E = \frac{6\pi}{80}\frac{e}{c}\left(\frac{\dot{M}}{V_\mathrm{w}}\right)^{1/2}\frac{\eta}{\Lambda} \frac{v_\mathrm{s}^3}{r} \propto t^{2j-3}
\end{equation}
or, if $\eta_\mathrm{B}$ is constant in time,
\begin{equation}
    \frac{dE}{dt} \propto t^{j-2}.
\end{equation}

It should be noted that, unlike for KOLM and KRAI turbulence, in this case the energy gain decreases with time faster than $\propto 1/t$ for $j<1$ implying that 
the maximum energy would depend on the time when particles are injected into the acceleration process. 
This is a very peculiar situation due to the fast decrease of the magnetic field with the distance from the origin and the rapid (linear) energy increase of the diffusion coefficient. Essentially this means that the instantaneous maximum energy would decrease with time and particles injected at later stages will not be accelerated to energies as high as those injected at the very beginning. Hence, although particles injected at the beginning will continue to gain energy the effective maximum energy will decrease with time. 

Integrating Eq.~\ref{eq:dEdtBohm} from the injection time $t_\mathrm{inj}$ to $t$ results in a dependence
\begin{equation}
    E_\mathrm{max} \propto \frac{1}{2-2j}\left(t_\mathrm{inj}^{2j-2} - t^{2j-2}\right)
\end{equation}
where for $j<1$ and $t_\mathrm{inj}<<t$ the maximum achievable energy of an injected particle would depend only on the time of injection. For particles injected at time $t_\mathrm{inj}<<t_\mathrm{TS}$ the maximum achievable energy can be estimated as 
\begin{multline}
    E_\mathrm{max} \simeq 153 \,\frac{j}{2-2j} \left[\frac{t_\mathrm{inj}}{t_\mathrm{TS}}\right]^{2j-2}\left[\frac{\eta}{0.1}\right]\left[\frac{\Lambda}{10}\right]^{-1}\left[\frac{\dot{M}}{2\times10^{-4}\,M_\odot/\mathrm{yr}}\right]^{1/2}\\
    \left[\frac{v_\mathrm{w}}{3\times10^8\,\mathrm{cm/s}}\right]^{-1/2} \left[\frac{v_\mathrm{TS}}{10^9\,\mathrm{cm/s}}\right]^{2}\,\mathrm{TeV}
\end{multline}
of for $j = 6/7$
\begin{multline}
    E_\mathrm{max} \simeq 456\, \left[\frac{t_\mathrm{inj}}{t_\mathrm{TS}}\right]^{-2/7}\left[\frac{\eta}{0.1}\right]\left[\frac{\Lambda}{10}\right]^{-1}\left[\frac{\dot{M}}{2\times10^{-4}\,M_\odot/\mathrm{yr}}\right]^{1/2}\\
    \left[\frac{v_\mathrm{w}}{3\times10^8\,\mathrm{cm/s}}\right]^{-1/2} \left[\frac{v_\mathrm{TS}}{10^9\,\mathrm{cm/s}}\right]^{2} \,\mathrm{TeV}
\end{multline}
Unfortunately, these expressions are not of much help in that they suggest that the particles that are accelerated to the highest energies are injected at times where the amount of mass processed by the shock is small. Moreover, if the resonant instability regime operates on top of the pre-generated turbulence, the energy gain at later stages could be actually controlled by the KRAI or KOLM diffusion allowing the maximum energy to keep growing. It is worth stressing that these estimates of the maximum energy apply to individual particles, while the maximum energy that appears in the spectrum of accelerated particles also reflects the amount of mass processed during the acceleration process. Only a numerical calculation such as the one discussed below can purposefully address this issue. It should be noted that these estimates differ considerably from the similar analysis adopted by \citet{2023MNRAS.519..136V} suggesting that typical SNRs evolving in the wind bubble of the stellar cluster could accelerate particles to $\sim4$~PeV under the assumption of Bohm diffusion. In addition to a different choice for $\eta_\mathrm{B}$ which they set to unity, a major difference in their treatment is the assumption of constant shock velocity, which essentially implies that the maximum energy would not depend on the injection time.

The situation is going to be somewhat different once the SNR shock reaches the termination shock and starts propagating in the hot bubble of the shocked wind. Assuming that the density and the magnetic field remain constant in the downstream of the termination shock, the energy gain will evolve with time as
\begin{equation}
    \label{eq:dEdtBohm}
    \frac{dE}{dt} \propto v_\mathrm{s}^3 \propto t^{3j-3}
\end{equation}
If the SNR is still in the free-expansion stage of its evolution, i.e. $j=2/3$ for expansion in a medium with constant density, the energy gain will decrease with time as $\propto 1/t$ and hence $E_\mathrm{max}$ would stay constant. On the other hand, as soon as the SNR transitions to the Sedov-Taylor stage, the shock starts slowing down faster, with $j=2/5$, and the energy gain starts decreasing as $\propto t^{-9/5}$ which leads to the saturation of the value of the effective maximum energy.
Other effects are also expected to play an important role. In particular, the interaction of the SNR shock with the termination shock results in formation of a reflected shock which after bouncing off the contact discontinuity inside the remnant would catch up with the forward shock accelerating it and leading to a slight increase of the maximum energy of accelerated particles \citep{2022ApJ...926..140S}. Another important effect is the weakening of the shock due to the high temperature of the medium which results in a softer spectrum of the accelerated particles \citep{2022A&A...661A.128D}. At this point a negligibly small fraction of the accelerated particles can reach very high energies. Both these effects are further discussed below in the context of numerical simulations.


So far we have discussed the effects of the excitation of the resonant streaming instability, but at fast SNR shocks a non-resonant instability is also expected to grow \citep{2004MNRAS.353..550B}, provided 
\begin{equation}
    \label{eq:bell}
    \frac{v_\mathrm{s}}{c} \frac{\eta}{\Lambda} v_\mathrm{s}^2 \rho > \frac{B^2}{4\pi}.
\end{equation}
For fast SNR shocks, the growth of the non-resonant instability is typically faster than the resonant branch. The condition above can be translated into the upper limit on the background magnetic field in the collective wind for the instability to grow:
\begin{multline}
     B < 1.92 \left[\frac{\eta}{0.1}\right]^{1/2}\left[\frac{\Lambda}{10}\right]^{-1/2} \left[\frac{\dot{M}}{2\times10^{-4}\,M_\odot/\mathrm{yr}}\right]^{1/2} \\
     \left[\frac{v_\mathrm{w}}{3\times10^8\,\mathrm{cm/s}}\right]^{-1/2}  \left[\frac{v_\mathrm{TS}}{10^9\,\mathrm{cm/s}}\right]^{3/2}\left[\frac{R_\mathrm{TS}}{20\,\mathrm{pc}}\right]^{-1} \left[\frac{t}{t_\mathrm{TS}}\right]^{(j-3)/2} \,\mu\mathrm{G}.
     \label{eq:Bnonres}
\end{multline}
Comparison with the magnetic fields expected from a $\sim 10\%$ conversion efficiency from the kinetic energy of the wind \cite[]{2023MNRAS.523.4015B}, immediately suggests that in order for this non-resonant instability to grow this efficiency must be appreciably smaller than $\sim 10\%$. If this is the case, then the magnetic field produced at saturation of the non-resonant instability is the same as given in the RHS of Eq. \ref{eq:Bnonres}, due to the way this instability works. Moreover, if the spectrum of accelerated particles is the canonical $p^{-4}$, then the diffusion coefficient is Bohm-like in the amplified magnetic field \citep{2020APh...12302492C,2021A&A...650A..62C} and it may easily be smaller (faster acceleration) than in the case of resonant instability considered above. A somewhat more stringent condition on the amplified magnetic field is also obtained based on the condition that the non-resonant instability grows fast enough in a given time $t$ \citep{2014MNRAS.437.2802S}. Since we will not consider the case of non-resonant instability below, we do not provide here additional details, which can be found in the articles above. 

Finally, magnetic field amplification can occur because of a sort of kinematic dynamo effect \citep{2012MNRAS.427.2308D,
2009ApJ...707.1541B}: vorticity can be induced in a plasma where density fluctuations are present, in the presence of a CR pressure gradient. The amplified magnetic field can then be estimated as \citep{2012MNRAS.427.2308D}:
\begin{equation}
\label{eq:Bdynamo}
    \frac{B^2}{4\pi}\approx \left( \frac{\delta \rho}{\rho}\right)^2 \eta^2 \rho v_s^2,
\end{equation}
and assuming that the instability drives toward density fluctuations of order unity, $\delta\rho/\rho\sim 1$, the equation above becomes $\frac{B^2}{4\pi}\approx \eta^2 \rho v_s^2$, which is typically $\sim 3-30$ times larger than the energy density potentially provided by the non-resonant instability. In the following we will consider Bohm diffusion in a magnetic field amplified in this way as the most optimistic case of particle acceleration.



\section{Numerical simulations}
\label{sec:numsim}
\subsection{Numeric setup}
Numerical simulations of particle acceleration at a SNR shock evolving in the super bubble created by the compact stellar cluster were performed using the \textbf{R}adiation \textbf{A}cceleration \textbf{T}ransport \textbf{Pa}rallel \textbf{C}ode (RATPaC). The code is designed for the time- and spatially dependent treatment of transport of cosmic rays in SNRs. Initially the code was relying on the assumption of the Bohm-like diffusion and analytic or pre-calculated numeric treatment of the hydrodynamic evolution \citep{2012A&A...541A.153T, 2013A&A...552A.102T}, while more recently the modules for solving the transport equation for magnetic turbulence \citep{2016A&A...593A..20B} and hydrodynamic simulations on-the-fly \citep{2018A&A...618A.155S, 2019A&A...627A.166B} were introduced. Below we summarize the the numerical setup used in this work.

\subsubsection{Hydrodynamics}

\begin{figure}
    \centering
    \includegraphics[width=\linewidth]{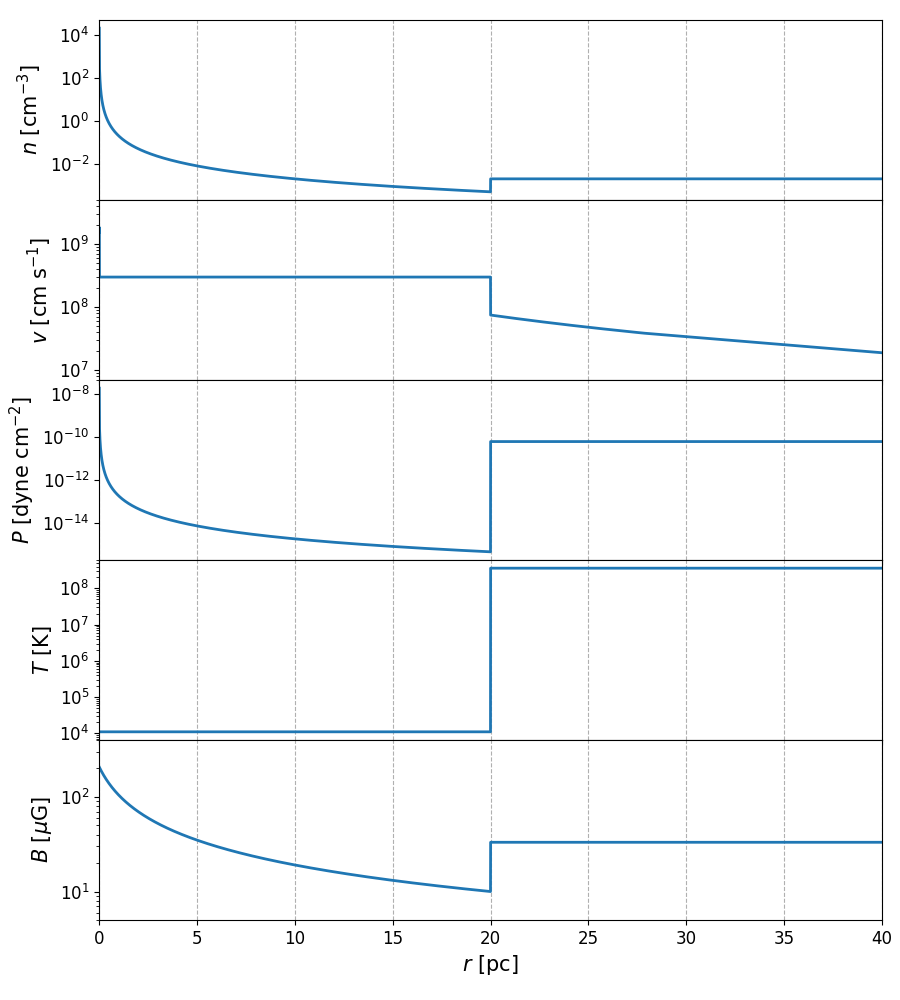}
    \caption{Magnetohydrodynamic radial profiles describing the star cluster bubble. From top to bottom: number density, plasma velocity, pressure, temperature, and magnetic field}
    \label{fig:hdprof}
\end{figure}

\begin{figure*}
    \centering
    \begin{subfigure}[b]{0.5\linewidth}
        \includegraphics[width=\linewidth]{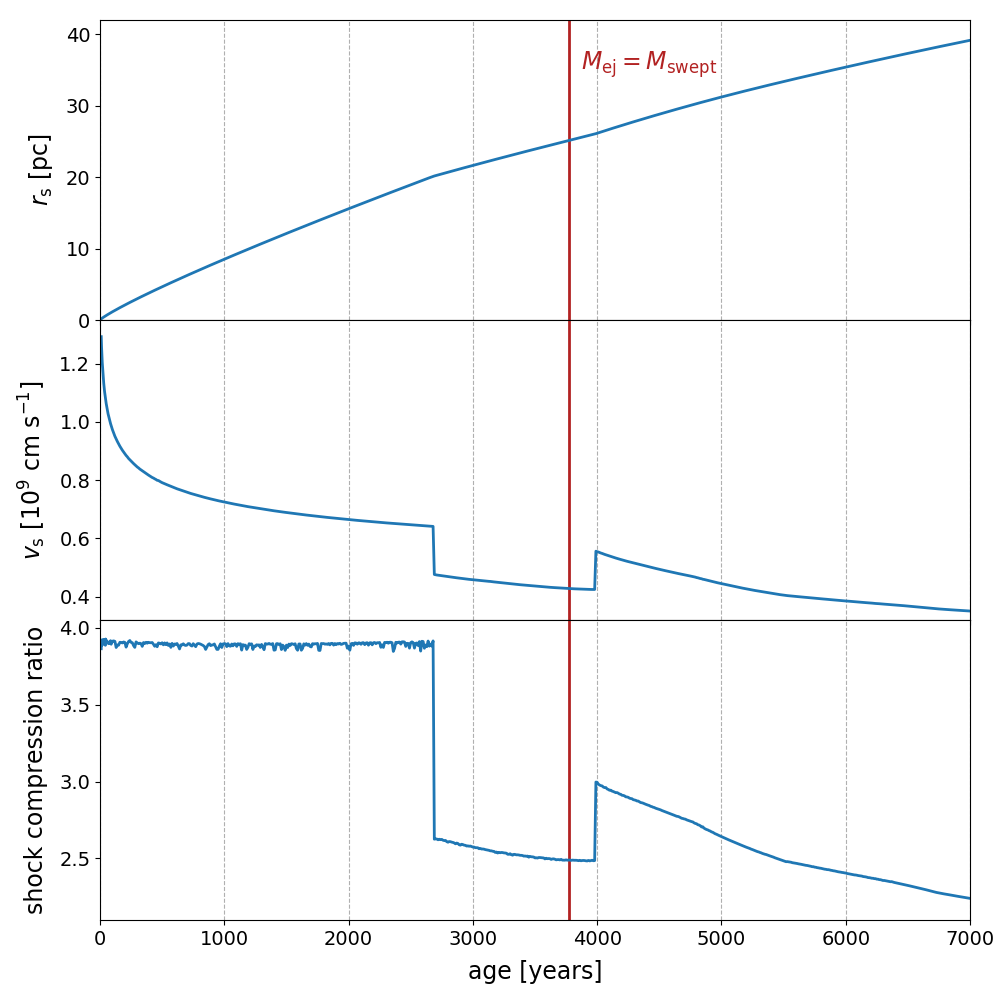}
    \end{subfigure}%
    ~
    \begin{subfigure}[b]{0.5\linewidth}
        \includegraphics[width=\linewidth]{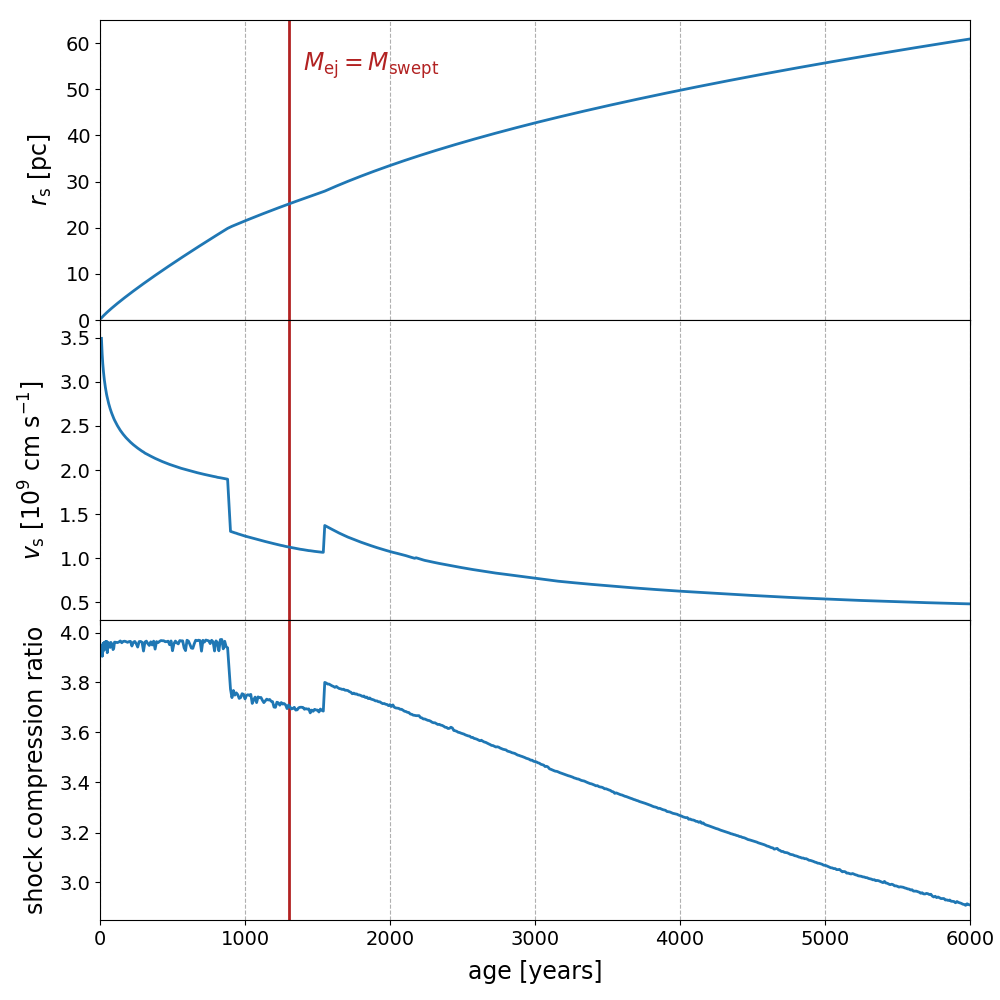}
    \end{subfigure}
    \caption{Evolution of the shock radius, velocity and compression ratio for the \textbf{classic} case on the left and \textbf{energetic} case on the right. The red vertical line indicates the time when the swept up by the shock mass becomes equal to the mass of the ejecta.}
    \label{fig:shockevo}
\end{figure*}

To describe the evolution of the SNR in the ambient medium we solve standard gas-dynamical equations ignoring the feedback of cosmic rays and magnetic field on the evolution.
\begin{align}
\frac{\partial }{\partial t}\left( \begin{array}{c}
                                    \rho\\
                                    \textbf{m}\\
                                    E
                                   \end{array}
 \right) + \nabla\left( \begin{array}{c}
                   \rho\textbf{v}\\
                   \textbf{mv} + P\textbf{I}\\
                   (E+P)\textbf{v} 
                  \end{array}
 \right)^T &= \left(\begin{array}{c}
                    0\\
                    0\\
                    0
                   \end{array}
 \right)\\
 \frac{\rho\textbf{v}^2}{2}+\frac{P}{\gamma-1}  &= E \text{,}
\end{align}
where $\rho$ is the density of the thermal gas, $\textbf{v}$ the plasma velocity, $\textbf{m}=\textbf{v}\rho$ the momentum density, $P$ the thermal pressure of the gas and $E$ the total energy density of the ideal gas with $\gamma=5/3$. This system of equations is solved under the assumption of spherical symmetry in 1-D using the \textsc{PLUTO} code \citep{2007ApJS..170..228M}.

The ejecta profile is initialized by a constant density,  $\rho_{\mathrm c}$, up to the radius $r_{\mathrm c}$, followed by a power-law distribution up to the ejecta-radius $R_{\mathrm{ej}}$,
\begin{align}
 \rho(r) &= \begin{dcases}
             \rho_{\mathrm c}, & r<r_{\mathrm c},\\
             \rho_{\mathrm c}\left(\frac{r}{r_{\mathrm c}}\right)^{-n}, & r_{\mathrm c}\leq r \leq R_{\mathrm{ej}},\\
            \end{dcases}
	\label{gasdyn}
\end{align}
beyond which the density profile is that of the star cluster super bubble described below. The exponent for the ejecta profile is set to $n=9$. The velocity of the ejecta is defined as
\begin{equation}
v_{\mathrm{ej}}(r) = \frac{r}{T_{\mathrm{SN}}},
\end{equation}
where $T_{\mathrm{SN}}=1$\,yr is the initial time set for hydrodynamic simulations. Defining the radius of the ejecta as multiple of $r_{\mathrm c}$, $R_{\mathrm{ej}} = xr_{\mathrm c}$ with $x = 3$, the initial conditions for simulations can be written as
\begin{align}
  r_c &=
  \left[\frac{10}{3}\frac{E_{\mathrm{ej}}}{M_{\mathrm{ej}}} \left(\frac{n-5}{n-3}\right) \left(\frac{1- \frac{3}{n} x^{3-n}}{1-\frac{5}{n} x^{5-n}} \right) \right]^{1/2} T_{\mathrm{SN}}, \\
  \rho_c &= \frac{M_{\mathrm{ej}}}{4\pi r_c^3}\frac{3(n-3)}{n} {\left(1- \frac{3}{n} x^{3-n}\right)^{-1}}, \\
  v_{\mathrm c} &= \frac{r_{\mathrm{c}}}{T_{\mathrm{SN}}}. 
\end{align}
In the following the ejecta mass is assumed to be $M_{\mathrm{ej}}=3\,M_\odot$ and for the explosion energy we consider two cases: \textbf{'classic'} with $E_\mathrm{ej} = 10^{51}$~erg and \textbf{'energetic'} with $E_\mathrm{ej} = 10^{52}$~erg.

We assume that the SNR explodes right on the edge of the star cluster core with $R_\mathrm{c} = 1$~pc. We redefine the spatial coordinate as $r = r - R_\mathrm{c}$ to have the SNR exploding at $r=0$ in our coordinate system. The medium created by the collective wind of the star cluster can be described by (Fig.~\ref{fig:hdprof})

\begin{align}
    \rho(r) &= \begin{cases}
            {\frac{\dot{M_\mathrm{w}}}{4\pi r^2 v_\mathrm{w}}}, & R_\mathrm{ej}<r\leq R_{\mathrm{TS}},\\
            4\rho(R_{\mathrm{TS}}), & r>R_{\mathrm{TS}},           
            \end{cases}\\
    v(r) &= \begin{cases}
            v_\mathrm{w}, &R_\mathrm{ej}<r\leq R_{\mathrm{TS}},\\
            0.25 v_\mathrm{w} (R_{\mathrm{TS}}/r)^2, & r>R_{\mathrm{TS}},
            \end{cases}\\
    p(r) &= \begin{cases}
            \rho(r) RT(r), &R_\mathrm{ej}<r\leq R_{\mathrm{TS}},\\
            3/4 \rho(R_{\mathrm{TS}})v_\mathrm{w}^2, & r>R_{\mathrm{TS}},
            \end{cases}\\
    T(r) &= \begin{cases}
            10^4 K, &R_\mathrm{ej}<r\leq R_{\mathrm{TS}},\\
            p(r)/R\rho(r),  & r>R_{\mathrm{TP}}.
            \end{cases}
	\label{WR_hydro}
\end{align}
as follows from Section~\ref{sec:coll_wind}. We set the total mass-loss rate to $\dot{M}=2\times10^{-4}\,M_\odot/\mathrm{yr}$ and wind velocity to $v_\mathrm{w}=3\times10^8$~cm/s. The location of the termination shock is assumed to be at $R_\mathrm{TS} = 20$~pc in the adopted coordinate system which corresponds to $21$~pc from the centre of the star cluster. Note, that here we ignore the outer part of the super bubble assuming that they are of no importance in the context of maximum reachable energies of particles accelerated at the SNR shock. The spatial grid used in the hydrodynamic simulations extends to $r=60\,$pc with $524288$ linearly spaced grid points. 
The evolution of the SNR shock throughout 7000 years in the classic case and 6000 years in the energetic case is shown on left and right panels of Fig.~\ref{fig:shockevo} respectively. The abrupt drop of the shock velocity at around 2600 years in the classic case and just before 1000 years in the energetic case is associated with the shock reaching the termination shock. At this moment, a reflected shock is formed, travelling into the SNR interior. At some point in time such a shock bounces off the contact discontinuity inside the remnant and starts propagating towards the forward shock. At around 4000 years in the classic case and 1500 years in the energetic case the reflected shock catches up with the forward shock giving it a boost. Once the shock starts propagating inside the downstream of the termination shock, the compression ratio decreases rapidly due to the high temperature of the medium there, leading to a small Mach number of the shock and a correspondingly steep spectrum of the accelerated particles. 

\subsubsection{Diffusion}

\begin{figure}
    \centering
    \includegraphics[width=\linewidth]{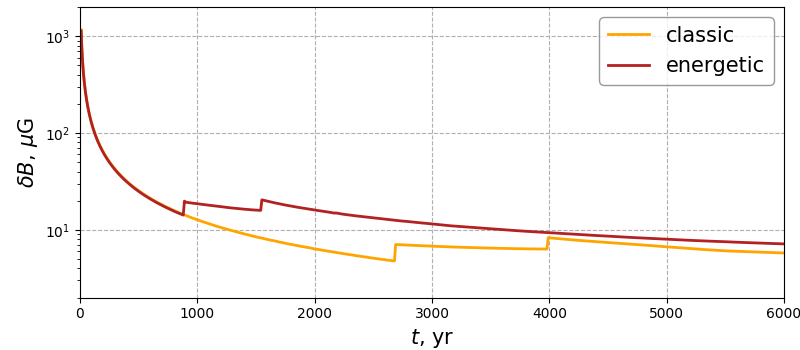}
    \caption{Magnetic field in the immediate upstream of the shock amplified through the kinetic dynamo effect for the classic and energetic hydrodynamic setups. In both cases $\eta = 0.1$ is assumed}
    \label{fig:deltaB}
\end{figure}

For the spatial diffusion we investigate different regimes as described in Section \ref{sec:Emax}. For pre-existing turbulence we use expressions for the KOLM and KRAI diffusion (Eqs.~\ref{eq:Dkolm} and \ref{eq:Dkrai}), applied to the whole simulation domain. The magnetic field is normalized to $10\,\mu$G in the wind at the location of the termination shock which corresponds to $210\,\mu$G at the edge of the core of the stellar cluster. The spatial dependence of the magnetic field follows Eq.~\ref{bturb}.

The possibility that magnetic field can be efficiently \textbf{self-generated} is studied assuming Bohm diffusion (BOHM; Eq.~\ref{eq:Dbohm} with $\eta_\mathrm{B} = 1$), and in order to maximize the effect on the maximum energy and provide an upper limit to the particle acceleration process, we consider the case of a magnetic field amplified through the kinematic dynamo effect following Eq.~\ref{eq:Bdynamo}. The particle injection efficiency $\eta$ is set at $10\%$ of the ram pressure for this calculation.
Note that in performed simulations the injection is prescribed as described in the next section which results in a time-dependent injection efficiency, but for the calculation of the amplified magnetic field we freeze at $\eta = 0.1$ to consider the most optimistic case that provides the highest value of the magnetic field. The amplified field acts only in the immediate upstream of the shock and exponentially decreases to the level of the background field on the length scale $\Delta l = 0.05 r_\mathrm{s}$. The choice of $5\%$ of the shock radius is consistent with the precursor length scale, $\sim D(E)/v_\mathrm{s}^2$, at $E\sim 1$~PeV. The background magnetic field in this scenario is assumed to be at the level of $10\%$ of the one assumed for the pre-existing turbulence, but this does not have important implications for the results. The amplified magnetic field $\delta B$ upstream of the shock is shown in Fig.~\ref{fig:deltaB} as a function of time calculated for the cases of standard and energetic supernova event. Initially $\delta B(t)$ is the same for both setups because $\delta B \propto v_\mathrm{s}/r_\mathrm{s}$ for the shock propagating in the wind zone of the bubble, but at later times some differences arise because in the energetic case the SNR shock reaches the termination shock faster than in the standard case. In both cases, the first sharp increase of $\delta B$ is associated with the SNR shock reaching the termination shock, where the upstream density increases by a factor of 4 and the second one is associated with the reflected shock catching up with the forward shock and giving it a boost. 
The Bohm-like diffusion is assumed to operate in the precursor, i.e. up to $1.05r_\mathrm{s}$, beyond which it transitions exponentially to the KOLM diffusion in the background field at two shock radii.

\subsubsection{Particle acceleration}

We use a kinetic approach to model the acceleration of CRs in the test-particle approximation. The time-dependent transport equation for the differential number density of cosmic rays, $N(p,r,t)$, \citep{Skilling.1975a} is described by the following equation:
\begin{equation}
\frac{\partial N}{\partial t}=\nabla(D\nabla N-\vec{u}N)-\frac{\partial}{\partial p}\left((N\dot{p})-\frac{\nabla \vec{u}}{3}Np\right)+Q,
\label{eq:tran}
\end{equation}
where $D$ denotes the spatial diffusion coefficient, $\vec{u}$ is the plasma velocity, $\dot{p}$ represents energy losses and $Q$ is the source of thermal particles. The energy-loss term is relevant only for electrons while here we only consider protons as accelerated particles. Equation (\ref{eq:tran}) is solved using the \textit{FiPy}-library \citep{2009CSE....11c...6G} in a frame co-moving with the shock. The radial coordinate is transformed according to $(x-1)=(x^*-1)^3$, where $x=r/r_{s}$. This transformation guarantees a very fine resolution close to the shock for an equidistant binning of $x^*$. The outer grid boundary extends to 65 shock radii upstream for $x^*_\mathrm{max}=5$. Thus, all accelerated particles can be kept in the simulation domain. 

Particle injection at the shock is described, for the sake of simplicity, through a thermal leakage model \citep{2005MNRAS.361..907B}, 
\begin{equation}
Q = \chi n_\mathrm{up}  (v_\mathrm{s} - {u_\mathrm{up}}) \delta(r-r_\mathrm{s}) \delta(p - p_{\mathrm{inj}}),
\end{equation}
where $\chi$ is the injection efficiency parameter,  $n_\mathrm{up}$ and $u_\mathrm{up}$ are the plasma number density and velocity in the upstream region, 
and $p_\mathrm{inj}= \xi p_\mathrm{th}$ is the injection momentum, defined as a multiple of the thermal momentum in the downstream plasma. In performed simulations we set $\xi = 4$ ensuring that the pressure of cosmic rays is always $<10\%$ of the ram pressure and hence cosmic rays are dynamically unimportant. The injection efficiency for the compression ratio of 4 is determined as
\begin{equation}
  \chi = \frac{4}{\sqrt{\pi}}\frac{\xi^3}{e^{\xi^2}}.
\end{equation}

\subsection{Results}

\begin{figure*}[h!]
    \centering
    \includegraphics[width=\linewidth]{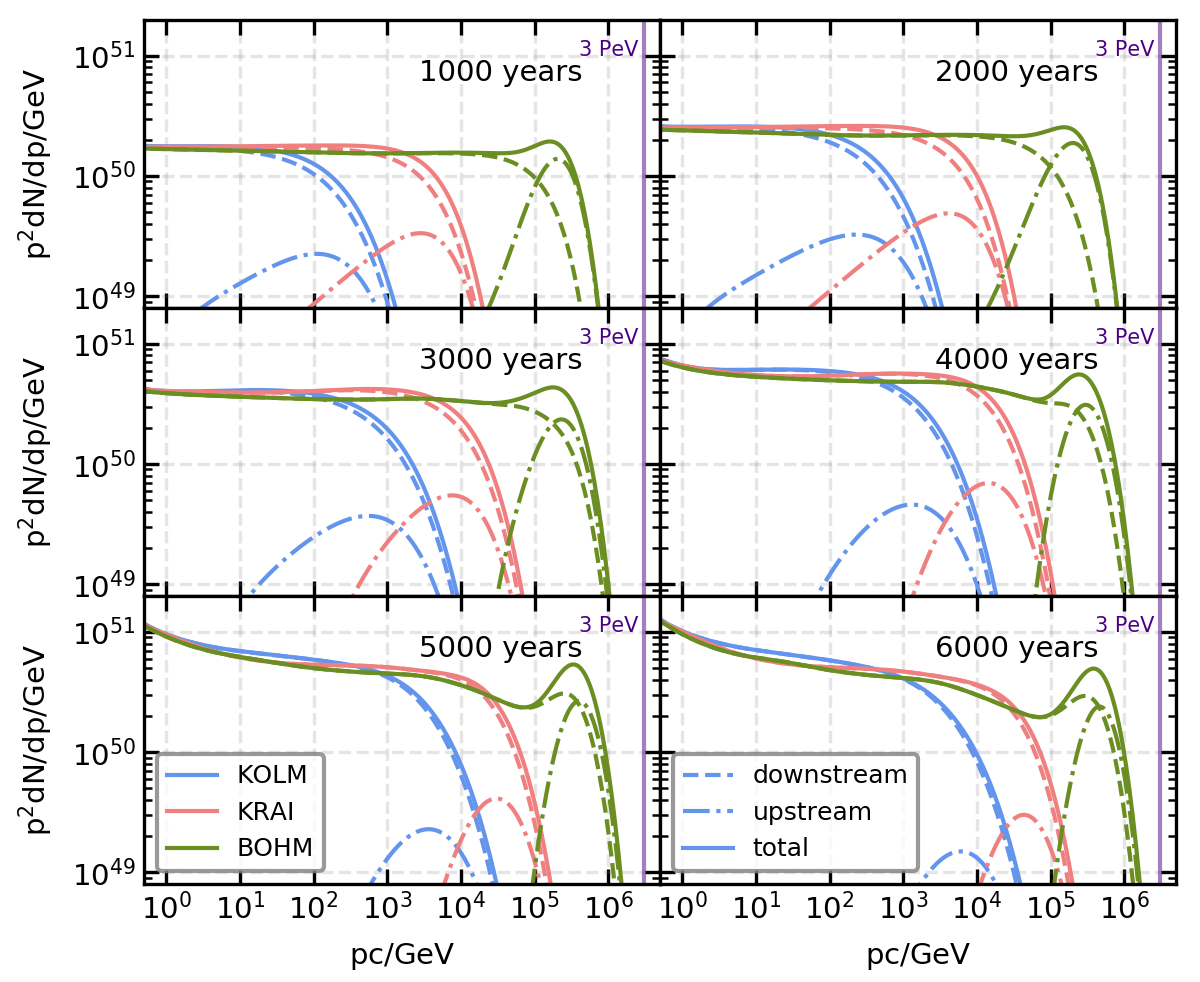}
    \caption{Classic case for different turbulence models and ages}
    \label{fig:prcl}
\end{figure*}

In Fig.~\ref{fig:prcl} we show the evolution of the proton spectrum for the case of a standard supernova explosion and different types of diffusion. Solid lines represent the total spectrum, while dashed and dash-dotted lines correspond to the downstream and upstream spectra respectively. Vertical purple lines indicate the 'knee' energy of 3 PeV. It can be seen that PeV energies are hard to reach in any of the considered scenarios. In the most optimistic case of Bohm diffusion the maximum energy reaches at most a couple of hundred TeV. At around 2600 years the SNR shock hits the termination shock and slows down considerably. Downstream of the termination shock the gas is hot and the Mach number of the SNR shock decreases significantly. The compression ratio of the shock drops to almost $2.5$ resulting in a very soft spectrum of freshly injected and accelerated particles with a spectral index of $\sim3$. Essentially, this means that the acceleration of freshly injected particles becomes irrelevant for the maximum energy at this stage. However, high-energy particles accelerated in earlier stages become compressed at the slowed down forward shock and may further gain some energy. 

Roughly $4000$ years after the explosion, the reflected shock reaches the forward shock slightly energizing it. However, this does not cause any visible effect of the spectrum of accelerated particles. With time particles injected after crossing the termination shock start dominating over the particles accelerated at earlier stages, which is evident from the soft spectrum at lower energies. It is worth to keep in mind that this conclusion depends on the adopted recipe for injection: for shocks with Mach number $\sim 2.5$ particle acceleration is expected to shut down \cite[]{Vink2014}, in which case this effect would disappear. 

The mass of the swept-up material becomes equal to the mass of the ejecta shortly before 4000 years, meaning that at $\sim$ 5000-6000 years the SNR is already deep in the Sedov-Taylor stage of the evolution and we should not expect further increase of the maximum energy.


\begin{figure*}
    \centering
    \includegraphics[width=\linewidth]{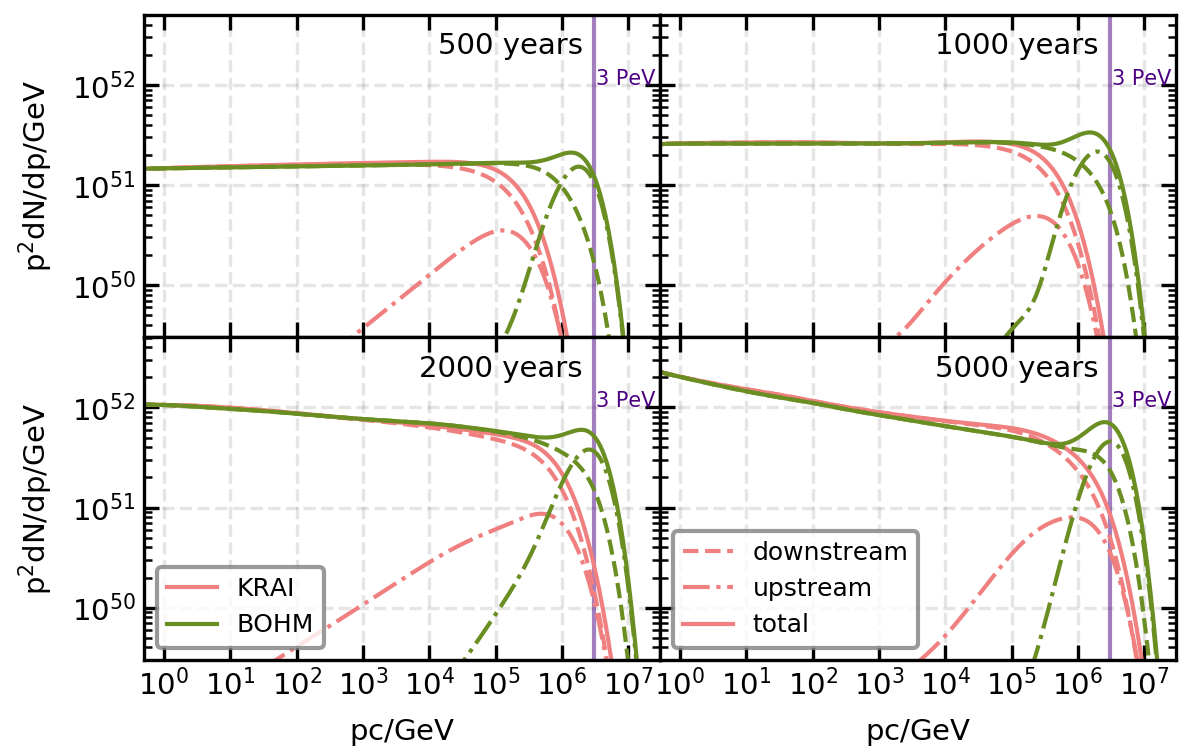}
    \caption{Energetic case for KRAI and BOHM turbulence models for different ages}
    \label{fig:energetic}
\end{figure*}


    


In the energetic case the situation is certainly more favourable. Higher shock velocity ensures acceleration of protons to a few hundred TeV even in the case of the pre-existing turbulence with Kraichnan diffusion coefficient. In Fig.~\ref{fig:energetic} the evolution in time of the total particle spectrum is shown for such energetic scenario, for the case of KRAI and BOHM diffusion. The only difference with respect to the models presented in Fig.~\ref{fig:prcl} is the explosion energy of the supernova. The forward shock reaches the termination shock shortly before 1000 years while the interaction with the reflected shock takes place at about 1500 years. The evolution of the spectra is similar to the classical case with the main difference that higher maximum energies could be reached. Also, as a result of a higher shock velocity the compression ratio of the shock stays high in the downstream of the termination shock ($\sim 3.7$ right after crossing the termination shock) for longer times and eventually falls down to $\sim3$ only $\sim 5000$ years after explosions, when the remnant is already very deep in the Sedov-Taylor stage. This increases the importance of particles injected after crossing the termination shock in the very-high-energy part of the spectrum and essentially prevents the formation of the pile-up feature in the downstream spectrum. It can be seen that in the case of the BOHM diffusion in the amplified field, as inferred for the kinetic dynamo effect, the spectrum of accelerated particles reaches PeV energies with a maximum energy saturating around the "knee"-energy of $\sim 3$~PeV. It should be understood however that this result can only be achieved for rare, very energetic supernova explosions, and adopting the most optimistic assumptions on particle diffusion.



\section{Summary and discussion}
\label{sec:summary}

We numerically investigated the acceleration of cosmic rays at the shock of a supernova exploded in the outskirts of a star cluster core, for a standard choice of the parameters of the explosion and for an energetic case. The shock expands in the bubble excavated by the collective wind of the cluster and its dynamics is modified by this environment. Our analysis shows that the acceleration history does not differ sensibly from the case of an individual core-collapse SNR in the stellar wind bubble of the progenitor star and essentially faces the same difficulties when it comes to the acceleration to PeV energies. When the diffusion coefficient is dominated by the pre-existing magnetic turbulence, presumably present in the super bubble due to the interaction of winds of individual stars, appears to be insufficient to warrant acceleration up to PeV energies. Only for energetic events with an explosion energy of $10^{52}$~erg the pre-existing turbulence is sufficient to accelerate particles to a few hundred TeV. The maximum energy for a standard SNR drops to values of a few tens of TeV. 

The general conclusion is that, similar to the case of isolated SNRs, acceleration to higher energies require self-generation of magnetic turbulence through streaming instabilities or other mechanisms. When the explosion occurs in a super-bubble, the conditions are adverse to the development of a non-resonant streaming instability, as previously pointed out by \cite{2021MNRAS.504.6096M, 2023MNRAS.523.4015B}, while resonant streaming instability seems insufficient to improve the odds of acceleration to PeV energies.


The most optimistic scenario that we consider is that of amplification of the background magnetic field through the kinetic dynamo effect, which could potentially provide the conditions for enhancing the magnetic field to larger values and lead to Bohm diffusion \citep{2012MNRAS.427.2308D,
2009ApJ...707.1541B}. In this borderline case, PeV energies can be reached only for very energetic, rare SNR explosions, not very different from the case of an isolated SNR.


\isrev{Finally, it should be noted that while this manuscript was in revision a new study was published proposing magnetic mirroring as an efficient mechanism to reach the level of Bohm diffusion in the presence of magnetic fluctuations at scales larger than the gyroradius \citep{2025MNRAS.539.1236B}. If this mechanism is indeed in place the maximum energy would be pushed to PeV for pre-generated turbulence that we consider in the collective wind of the young compact stellar cluster. However, a similar environment is also expected for isolated young core-collapse SNRs and obsevations do not show any evidence of acceleration to PeV energies in SNRs so far.}

\begin{acknowledgements}
IS acknowledges funding from Comunidad de Madrid through the Atracción de Talento “César Nombela” grant with reference number 2023-T1/TEC-29126. The work of PB has been partially funded by the European Union - Next Generation EU, through PRIN-MUR 2022TJW4EJ and by the European Union - NextGenerationEU under the MUR National Innovation Ecosystem grant  ECS00000041 - VITALITY/ASTRA - CUP D13C21000430001, and by the research project TAsP
(Theoretical Astroparticle Physics) funded by the Istituto Nazionale di Fisica Nucleare (INFN).
\end{acknowledgements}

%
%

\bibliographystyle{aa} 
\bibliography{bibliography.bib}

\end{document}